\documentstyle[twoside,fleqn,espcrc2,epsf]{article}

\def\mtau{m_\tau}
\def\mtau2{m^2_{\tau}}

\def\e{ \rm{e}}     
\def\ee{ \mbox{\rm e}^+ \mbox{\rm e}^-}
\def\tt{ \tau^+\tau^-}
\def\mm{ \mu^+\mu^-}
\def\qq{ \mbox{\rm q} \bar{\mbox{\rm q}}}

\def\eeha{\ee \rightarrow \qq}
\def\ggee{\ee \rightarrow (\ee)\ee}

\newcommand{\aone}{\mbox{\rm a}_1}
\def\a1nu{\aone\nu}

\def\s2thw{\sin^{2}\theta^{\mbox{\scriptsize lept}}_{\mbox{\scriptsize eff}}}

\newcommand{\TMU}{\tau \rightarrow \mu \nu \bar{\nu}}
\newcommand{\TEL}{\tau \rightarrow \mbox{\rm e} \nu \bar{\nu}}

\newcommand{\AmS}{{\protect\the\textfont2
  A\kern-.1667em\lower.5ex\hbox{M}\kern-.125emS}}

\title {Summary on Tau Leptonic Branching Ratios and Universality} 

\author{B. Stugu\address{Department of Physics, University of Bergen \\
All\'{e}gaten 55, N-5007 Bergen} }
\begin{document}

\begin{abstract}
The large samples of $\tau$ decays available from CLEO
and the four LEP experiment have resulted in new, precise
measurements of the leptonic branching ratios of the $\tau$.
The experimental techniques to obtain these results are
reviewed with special emphasis on the DELPHI measurement.
World averages 
are found to be $B(\TEL) = (17.81 \pm 0.06 \%)$
and $B(\TMU) = (17.36 \pm 0.06 \%)$. These results are 
consistent with universality in the charged
current couplings to a precision of about 0.25 \%. 
The branching ratio measurements can
also be used to constrain the "low energy parameter" $\eta$.
It is shown that the sensitivity to $\eta$ depends on
details of the momentum acceptance for muon identification in the
different experiments. Assuming universality
in the couplings, the estimate $\eta = 0.012 \pm 0.024$ is
obtained. 
\end{abstract}

\maketitle

\section{ INTRODUCTION}
Universality in the couplings of the three lepton families to the gauge bosons  
is a fundamental assumption of the standard model.  As any observed deviation
from universality would imply the presence of physics beyond the standard
model, it is important to test this assumption as precicely as possible.
Due to the accurate predictions of the 
$\tau$ leptonic decays rates, given by\cite{SIRLIN}: 
\begin{equation}
\label{eq1}
 \Gamma(\tau \rightarrow l \nu_{\tau} \overline{\nu}_{l}) =
\frac{B_l}{\tau_{\tau}} =
 \frac{G_{l \tau}^{2}m_{\tau}^{5}}{192 \pi^3}f(x_{l}^2) r_{RC},
\end{equation}
measurements of  the branching fractions
$B_{\e} = B(\TEL)$ and $B_{\mu} = B(\TMU)$ will, together with the
$\tau$ and muon masses and lifetimes, give unambigious tests
of universality in the couplings of all three lepton families to the
$W$-boson. 
In eq. \ref{eq1},  $\tau_{\tau}$ is the lifetime of the
$\tau$-lepton, and the function $f(x_{l})$, with $x_{l}= \frac{m_l}{m\tau}$,
is a phase space factor with $f(x_{\e}) = 1$ and $f(x_{\mu}) = 0.9726$.
The factor $r_{RC}$ accounts for radiative corrections, which, for
practical purposes can be taken to be of equal magnitude for the
rates $\Gamma_e$ and $\Gamma_{\mu}$.
\par
In the
standard model, the coupling $G_{l \tau}$ is given by
\begin{equation}
\label{eq2}
  G_{l \tau}^2 = \frac{g_l^2 g_\tau^2}{32 m_W^4} 
\end{equation}
and equals the Fermi coupling constant if universality holds. As
seen from equation \ref{eq1}, the ratio
\begin{equation}
\label{eq3}
\frac{B_{\mu}} {B_e} =
\frac{g_{\mu}^{2}} {g_{e}^{2}} \cdot
\frac{f(x_{\mu)}^2} {f(x_{\e}^2)} 
\end{equation}
eliminates $g_{\tau}$, giving a direct  comparison between 
$g_{e}$ and $g_{\mu}$.
\par
A comparison of $g_{\tau}$ to the couplings to the two lighter
leptons requires $\tau$ and muon masses and lifetimes. Using the
analogue of equation \ref{eq1} for muon decays, $g_{e}$ can be 
eliminated to give a test of $\tau$-$\mu$ universality:
\begin{equation}
\label{eq4}
   B_e =
\frac{g_{\tau}^{2}} {g_{\mu}^{2}} \cdot
    \left[ \frac{ f( x_{\e}^2) r_{RC}^\tau}
               { f ( x_{\mu \e}^2 ) r_{RC}^\mu}
      \right]
\frac{m_{\tau}^{5}}{m_{\mu}^{5}} \frac{1}{\tau_{\mu}} \cdot
     \tau_{\tau},
\end{equation}
and finally $\tau$-e universality is tested in confronting:
\begin{equation}
\label{eq5}
   B_{\mu} =
\frac{g_{\tau}^{2}} {g_{\e}^{2}} \cdot
    \left[ \frac{ f ( x_{\mu}^2 ) r_{RC}^\tau}
               { f ( x_{\mu \e}^2 ) r_{RC}^\mu}
      \right]
\frac{m_{\tau}^{5}}{m_{\mu}^{5}} \frac{1}{\tau_{\mu}} \cdot
     \tau_{\tau},
\end{equation}
with measurement. Here 
$x_{\mu \e} = \frac{m_{\e}}{m_{\mu}}$.
\par
This talk reviews the newest measurements
of $B_{e}$ and $B_{\mu}$ and uses the results to give estimates of
the ratios between the couplings of the weak charged current to
the different leptons. Special attention will be made to the
new DELPHI result, the only new measurement  which has not been discussed in
a separate presentation at this conference.
Brief mention will also be made
of other ways to test universality of the charged current, comparing
the precision of these to the precision obtained with leptonic $\tau$
decays. Finally a remark will be made concerning the sensitivity of
the branching ratio measurements
to the "low energy parameter", $\eta$, and an estimate of this
parameter will be made.

\section{THE MEASUREMENTS}
 The world averages for $B_e$ and $B_{\mu}$ are dominated by the
results from CLEO and the four LEP experiments.   In the following
attention will be given to the differences in the techniques used
to extract the branching ratios. These differences are a concequence of
the much smaller centre of mass energy at CESR compared to
that at LEP.  The current
CLEO and ALEPH best estimates are published \cite{CLEO}, \cite{ALEPH}.
while L3, OPAL and DELPHI values are still preliminary.
With the separate presentation of the new OPAL 
$B_e$ measurement at this workshop \cite{OPAL1}, all measurements 
have been presented in this
workshop series \cite{OLDTAU}, except the new DELPHI results \cite{VANC},
and the OPAL $B_{\mu}$ measurement which was available already last 
year \cite{OPAL2}. 
Particular attention is thus given to 
the DELPHI measurements here (sect. 2.3).
\subsection{ The CLEO measurement}
In $\ee$ collisions at the $\Upsilon$(4s) energy, it is difficult
to distinguish a produced 
$\tt$ pair from a $\qq$ pair unless requirements are made on the
decay of at least one of the two $\tau$-leptons produced.   The 
CLEO measurement \cite{CLEO}, thus makes a selection of 
1-prong 1-prong $\tt$ decays as a starting point for their
analysis. The data are divided into classes $ab$, where
the first $\tau$ has the decay mode $ \tau_{1} \rightarrow a$,  and 
the second $\tau$ decays as  $ \tau_{2} \rightarrow b$. It is noted that the
number of events, $n$, in class $ab$ is given by the product:
\begin{equation}
\label{eq6}
 B_a \times B_b = \frac{ n \times ( 1 - f)}
      {\varepsilon \times N_{\tau \tau} \times ( 2- \delta_{ab} )}
\end{equation}
Here, $f$  is the fractional background, 
 $\varepsilon$ is the efficiency  of selection and
$\delta_{ab}$ is Kronecker delta.
The impressive number of around three million $\tau$ pairs produced, 
$N_{\tau \tau}$, is determined from
the theoretical cross section and the integrated luminosity. The analysis
 selects $\tau$ decays with at most one $\pi^0$ present and uses the CLEO value
for $B(\tau^{\pm} \rightarrow \pi^{\pm} \pi^0)$ as input to perform a
simultaneous determination of $B_{e}$, $B_{\mu}$ and 
$B_{h} = B(\tau \rightarrow \pi (K) \nu)$, as well as the
ratios $B_\mu/B_{\rm{e}}$ and $B_h/B_{\rm{e}}$. Due to the very large
sample of $\tau$-leptons, the analysis obtains world record statistical
precision on $B_{e}$ and $B_{\mu}$.  The largest source
of uncertainty to the measurements is in $N_{\tau \tau}$, which has a relative
precision of about 1.4\%, contributing to the systematic uncertainty in
the branching ratio estimates with about 0.7\% (relative). This causes
the overall uncertainty of the branching ratio measurements to be 
dominated by  their systematic errors.
Much of this is in common  and cancels in the ratios.
\subsection{ Measurements at  LEP  }
The high multiplicity of quark jets at LEP makes a generic
$\tt$ preselection possible, and a simple
multiplicity requrement (e.g. asking the number of charged tracks
to be below 7) rejects most $\qq$ events.  The remaining sample
of events is mainly composed of $\ee$, $\mm$ and $\tt$ pairs, and appropriately
exploiting the presence of at least two unseen neutrinos in the two 
$\tau$ decays permits the selection of $\tt$ pairs with high efficiency
and purity, irrespective of the decay modes.
Hence, the following expression can be used
for computing the branching ratio:
\begin{equation}
\label{eq7}
  B(\tau\rightarrow l \nu\bar{\nu}) = \frac{N_l} {N_{\tau}} \cdot
				      \frac{1-b_l} {1-b_{\tau}} \cdot 
				      \frac{\epsilon_\tau}
				      {\epsilon_l } ,
\end{equation}
 where $N_{\tau}$ number of $\tau$ decays preselected , 
  $N_l$ is the number of identified leptons out of this sample,  
$\epsilon_\tau$, $\epsilon_l$ are $\tau$ and lepton selection efficiencies; and
 $ b_{\tau}$ , $b_l$ are background fractions in $\tau$ and lepton samples.
With a completely unbiased $\tt$ preselection, the value
of $\epsilon_\tau$ would be irrelevant,  and the lepton identifiction
efficiency,  $\epsilon_l^{id} $, as computed with respect to the
selected sample of tau pairs could have replaced the total efficiency
$\epsilon_l$. However, the effect of a possible bias 
in the preselection precedure is not negligible and has to be evaluated. 
It is appropriate to factorize the identification efficiency 
into $\epsilon_l = \epsilon_{\tau}^l \times \epsilon_l^{id}$, where
$\epsilon_{\tau}^l $ is the efficiency of the $\tt$ preselection procedure for 
the decay mode $\tau \rightarrow   l \nu \overline{\nu} $.  Then the
the effect of the preselection requirements on the
bias factor, $\beta_l = \epsilon_{\tau} / \epsilon_{\tau}^l $ should be
evaluated as a source of systematic uncertainty. The
systematic uncertainties in the LEP measurement stem from
the precision of cross checks between data and simulation, and are thus
statistics driven. Apart from the  common uncertainty in the background
of the preselection sample, the systematic uncertainties in 
$B_{\e}$ and $B_{\mu}$ are mainly uncorrelated. However, as
the branching ratios are derived from a common sample of $\tt$ pairs,
there is statistical anticorrelation between the results
obtained in a given experiment. 
\subsection{  The DELPHI measurement }
This is a brief account of the analysis described in \cite{VANC}.
Correct assignment of the charged and neutral particles to a specific $\tau$
is ensured by dividing the event  into two hemispheres
by a plane perpendicular to the event thrust axis. 
The preselection of $\tt$ pairs is restricted to events where at least
one of the two leading charged tracks in each hemisphere have a polar 
angle between 43 and 137 degrees. 
Some additional restrictions
to fiducial volume are required depending on channel to be identified.
To reject $\eeha$ events, the charged track multiplicity in the event
is restricted to be between 2 and 6. Events from two photon interactions
are rejected by asking a visible energy of at least 0.175 times
the centre of mass energy. Furthermore it is required that the isolation
angle between tracks in different hemispheres should be larger than 160 degrees.
For events with two charged particles it is required that the missing
transverse momentum in the event should be larger than 0.4 GeV/c, and
that the acollinearity should be larger than 0.5 degrees.
\par
\begin{figure}
\begin{center}
\mbox{\epsfxsize 8.0cm\epsfbox{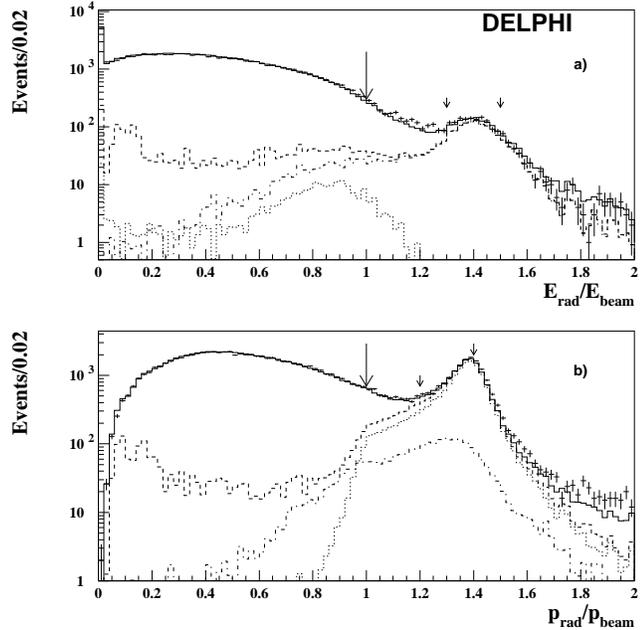}}
\end{center}
\vspace{-2.0cm}
\caption{ Distribution of the variables  a) $E_{rad}$ and  b) $p_{rad}$, 
designed to reject $\mm$ and $\ee$ pairs.} 
\label{fig1}
\end{figure}

After this,  significant amounts of $\ee$ and $\mm$ 
pairs are still present in the sample. These are dealt with by
exploiting the fact that much of the energy in the  $\tt$ events is
not seen due to the neutrinos. The variables
 $p_{rad} = \sqrt{p_1^2+p_2^2} $ and
$E_{rad} = \sqrt{E_1^2+E_2^2} $ are required to be smaller than 
the beam momentum and beam energy respectively. Fig. \ref{fig1} shows 
the distributions in these variable. Backgrounds are measured from
data by extrapolation from regions in the cut variables which
are dominated by a particular background type.  
\par
Having adjusted the backgrounds, remaining discrepancies between data
and simulation are assumed to affect the efficiency of the $\tt$ 
preselection, and possibly affect the bias factor, $\beta_l$.
The dependence of $\beta_l$  on a given selection variable is
determined by varying a cut around its chosen value, and a
systematic uncertainty is assigned based on the level of discrepancy 
observed comparing the number of events rejected in the data to
the corresponding number from simulation.
\par
\begin{figure}
\label{fig2}
\begin{center}
\mbox{\epsfxsize 6.8cm\epsfbox{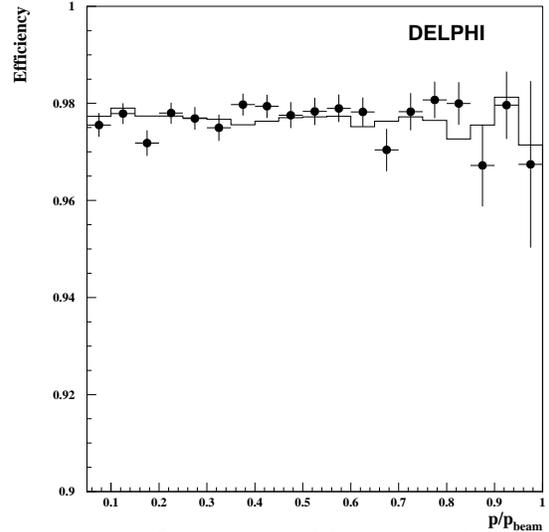}}
\end{center}
\vspace{-2.8cm}
\caption{ Comparison of data (points) and simulation 
(solid line) for the estimate of the efficiency of the .OR.
between the muon chamber requirement and the HCAL requirement.}
\end{figure}

The identification of muons is  done by requiring hits in the
muon chambers, or alternatively, requiring a significant
energy deposition in the outermost layer of the hadron calorimeter.
The efficiency of these requirements  are measured with respect
to eachother. Adjustments of the simulation efficiencies were
found to be necessary. After these adjustments, good data-simulation
agreement is found in the estimated efficiency of the combined
requirement, as shown in fig. \ref{fig2}.    The actual value for
the efficiency obtained by this procedure
is only valid for muons penetrating the whole detector, and
dimuon events were used to verify the overall efficiency. 
\par
\begin{figure}
\label{fig3}
\vspace{-2.5cm}
\begin{center}
\mbox{\epsfxsize 6.8cm\epsfbox{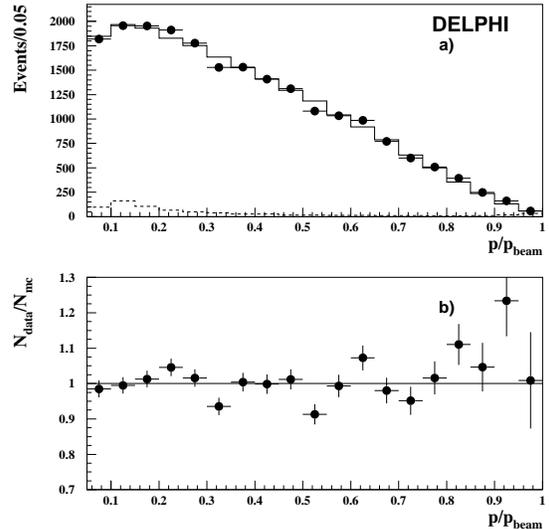}}
\end{center}
\vspace{-2cm}
\caption
{   a) Momentum distribution for identified muons  b) the
ratio between data and expectation from simulation}
\end{figure}

In order to
reduce backgrounds from hadrons further, it was required that the
average response per layer in the hadron calorimeter was compatible
with a minimum ionizing particle. Further supression of hadrons with
the presence of a $\pi^0$ was ensured by limiting the total
electromagnetic energy deposited in an 18 degree cone around the
charged  particle to 2 GeV. 
Selecting muons using tight requirements
on the muon chamber response  gave a very clean sample which
was used for  direct measurement of the
efficiency of all background suppression requirements. The levels of
the remaining backgrounds were verified by lifting one cut
to get a sample enhanced with a particular background and comparing
the effect of all other requirements on the data with the simulation
result. Furthermore the muon momentum distribution was studied with
and without a specific requirement to check that the data behaved
as the simulation. The final muon momentum distribution obtained 
is shown in fig. \ref{fig3}.
\par
\begin{figure}[tbph]
\label{fig4}
\begin{center}
\mbox{\epsfxsize 6.5cm\epsfbox{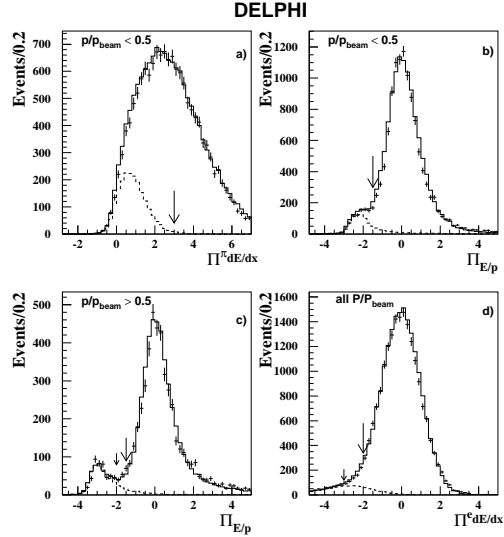}}
\end{center}
\vspace{-2.6cm}
\caption{ Variables for electron identification.
Large arrows show the cut values for identification.
Regions to the left of the small arrows are used to check the
background levels.}
\end{figure}

In order to identify electrons, pull variables defined from  the
energy loss in the TPC ($\Pi_{dE/dx}^{\e (\pi)}$, where the
superscript refers to the particle hypothesis),  and from the ratio 
between the electromagnetic
energy  and the particle momentum ($\Pi_{E/p}$)were defined.  Again the 
redundancy of
the two requirements could be used to check the efficiencies of these
requirements, though only for momenta between $0.05 \times p_{beam}$
and $0.5 \times p_{beam}$. In this region, requiring either the energy loss
should be incompatible with that expected for a pion, or a value of
$\Pi_{E/p}$ compatible with that expected for an electron ensured a
high, even and well controlled efficiency over this momentum range. 
for higher momenta, the efficiency requirement $\Pi_{E/p}> -1.5$ was 
checked with a sample of Bhabhas. Finally, all electron candidates
should have an energy loss compatible with that of an electron,
by requiring $\Pi_{dE/dx}^{\e }$ larger than -2.
Fig \ref{fig4} shows the distributions of the relevant identification
variables. 
\par
\begin{figure}[tbph]
\begin{center}
\mbox{\epsfxsize 6.5cm\epsfbox{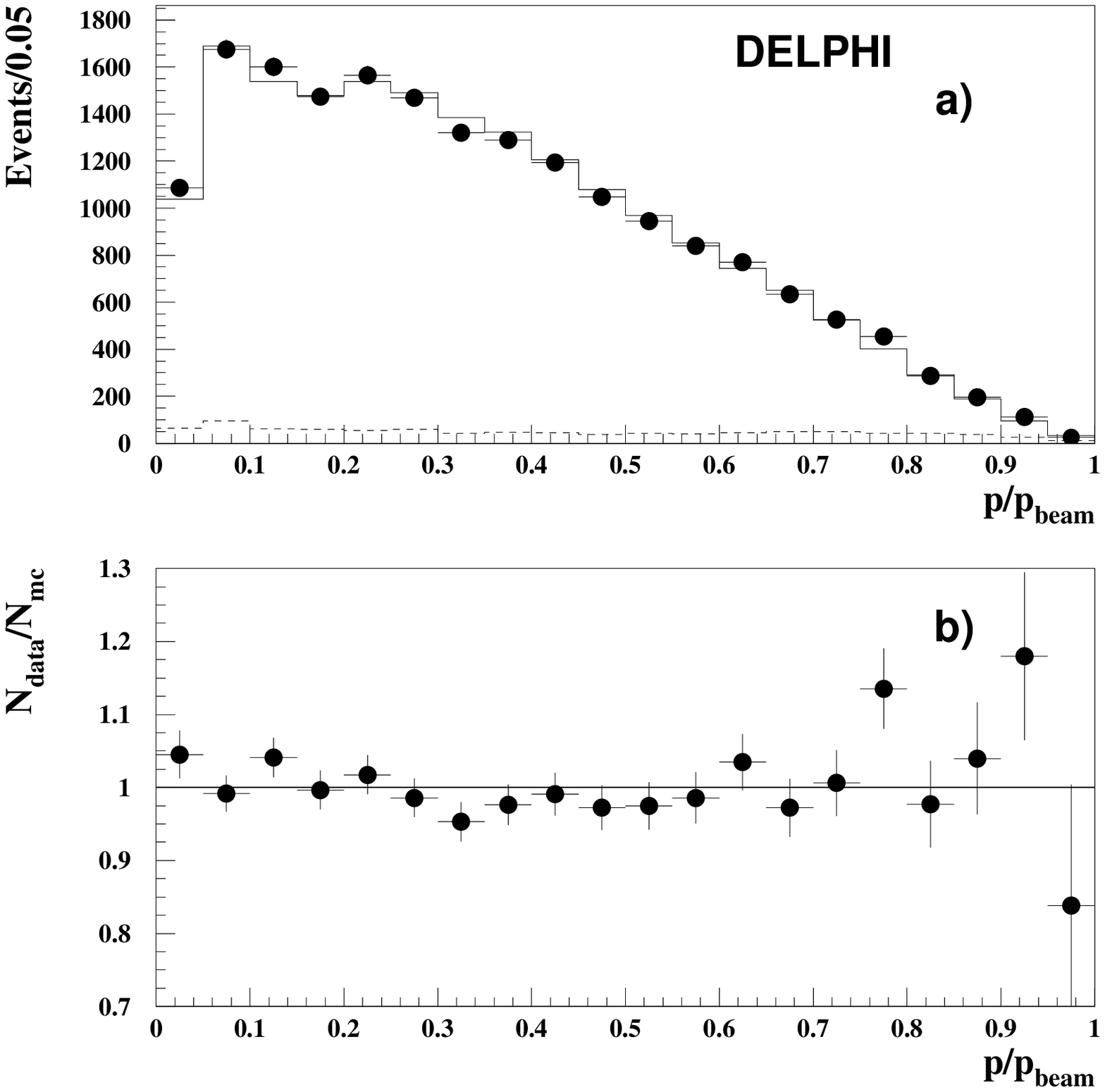}}
\end{center}
\vspace{-2.0cm}
\caption
{   a) Momentum distribution for identified electrons  b) the
ratio between data and expectation from simulation.}
\label{fig5}
\end{figure}

Much of the remaining background from hadrons was rejected
by requiring that no energy should be deposited beyond the first
layer of the hadron calorimeter, and  tau decays with the presence
of pions were rejected by vetoing candidates where the total
 electromagnetic 
energy deposition in a cone around the track exceeded 3 GeV.
 Here, neutral energy which
could stem from bremsstrahlung of the electron, was excluded from the sum.
The efficiency of all requirements  except the 
$\Pi_{dE/dx}^{\e }$ requirement, could be well measured by choosing
a clean sample of electrons, requiring $\Pi_{dE/dx}^{\e }> 0$. The
efficiency of the $\Pi_{dE/dx}^{\e } > -2.$ requirement was measured
using bhabhas. Finally, to reject $\ggee$ further, events  where
the momenta of both leading tracks were below $0.2 \times p_{beam}$,
and compatible with electrons were discarded from the sample. As
fig \ref{fig5} shows, the resulting momentum distribution of the 
final sample of electron candidates is well described by simulation.
\par
Table \ref{table1} summarizes the numbers entering the computation 
of the branching ratios for 93-95 data. For the $B_e$ measurement, 
the largest source of systematic errors is due to the uncertainty
in the identification efficiency estimate, while the systematic
error for the $B_{\mu}$ measurement is dominated by the uncertainty in
the bias factor.  The results from this analysis of 93-95 data
are combined with the previously published DELPHI result using
data from 1991 and 1992 to give the final DELPHI estimates of
$B_e$ and $B_{\mu}$ (shown in table \ref{table1}).
\begin{table*}[hbt]
\setlength{\tabcolsep}{1.5pc}
\label{table1}
\caption{Number of candidates, efficiencies and backgrounds
for the DELPHI 93-95 analysis and resulting $B_e$ and $B_{\mu}$ estimates.}
\begin{tabular*}{\textwidth}{@{}l@{\extracolsep{\fill}}rr}
\hline
Channel & $\TMU$ & $\TEL$ \\ \hline
 $N_{\tt}$  & $68655$ & $68668 $ \\
\hline
 $\epsilon_\tau$ (\%) & $52.57 \pm 0.04$ & $50.87 \pm 0.04 $  \\
  $ b_{\tau} $ & $ 3.09 \pm 0.11$ & $3.05 \pm 0.11 $  \\
\hline
$ N_l $ & 21040 & 18273 \\
 $\epsilon_l$ (\%) & $46.12 \pm 0.11 $ & $ 36.79 \pm 0.14 $ \\
$ b_l$ &  $ 3.65 \pm 0.16 $ & $ 5.23 \pm 0.30 $ \\
\hline
Branching ratio, 93-95  (\%) & $17.37 \pm 0.11_{stat} \pm 0.07_{sys} $ &
                               $17.98 \pm 0.12_{stat} \pm 0.09_{sys} $ \\
\hline
Branching ratio, 91-95  (\%) & $17.32 \pm 0.10_{stat} \pm 0.07_{sys} $ &
                               $17.92 \pm 0.11_{stat} \pm 0.10_{sys} $ \\
\hline
\end{tabular*}
\end{table*}
\section{ SUMMARY OF RESULTS AND UNIVERSALITY TESTS }
TAU98 has three  updated values  for $B_{\e} $ and $B_\mu$
compared to PDG98 \cite{PDG98} where  
nine measurements are used in each mode to compute averages. Hence
the number of measurements per mode is still nine. Of these, 
CLEO and the four LEP experiments have similar precisions
and combined they carry about 96.5 \% of the weights for both modes.
Almost the full dataset of the LEP experiments is now analyzed, with
the exception of ALEPH, where the analysis of only about half 
the LEP-I data is completed, and L3, where 1995 data are not included
in their measurements.  The measurements are summarized in
figs. 6 and 7.
\begin{figure}
\label{fig6}
\begin{center}
\mbox{\epsfxsize 8cm \epsfbox{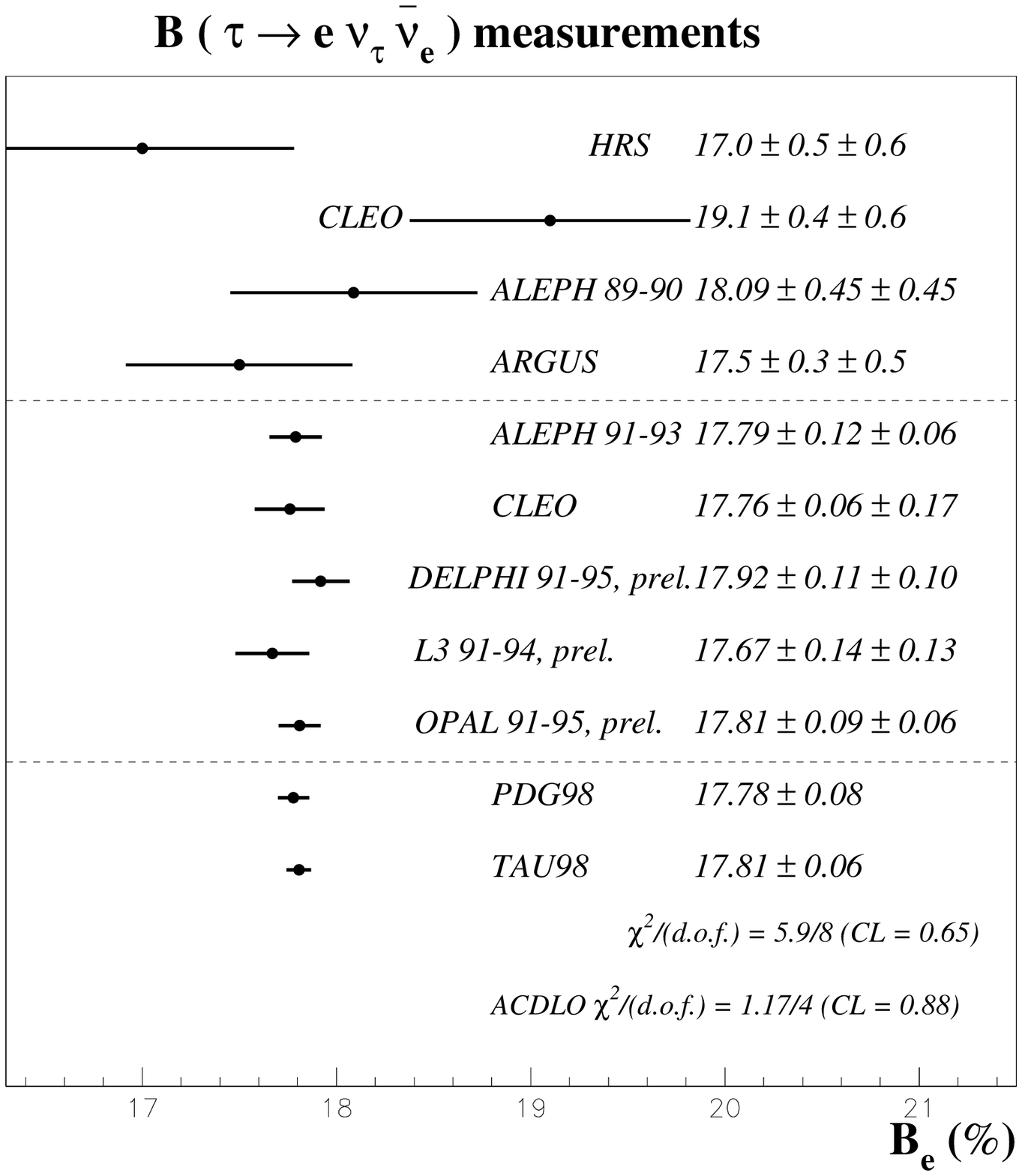}}
\end{center}
\vspace{-2.0cm}
\caption{ $B(\TEL)$ measurements for the world average.}
\end{figure}
\begin{figure}[t]
\label{fig7}
\begin{center}
\mbox{\epsfxsize 8cm \epsfbox{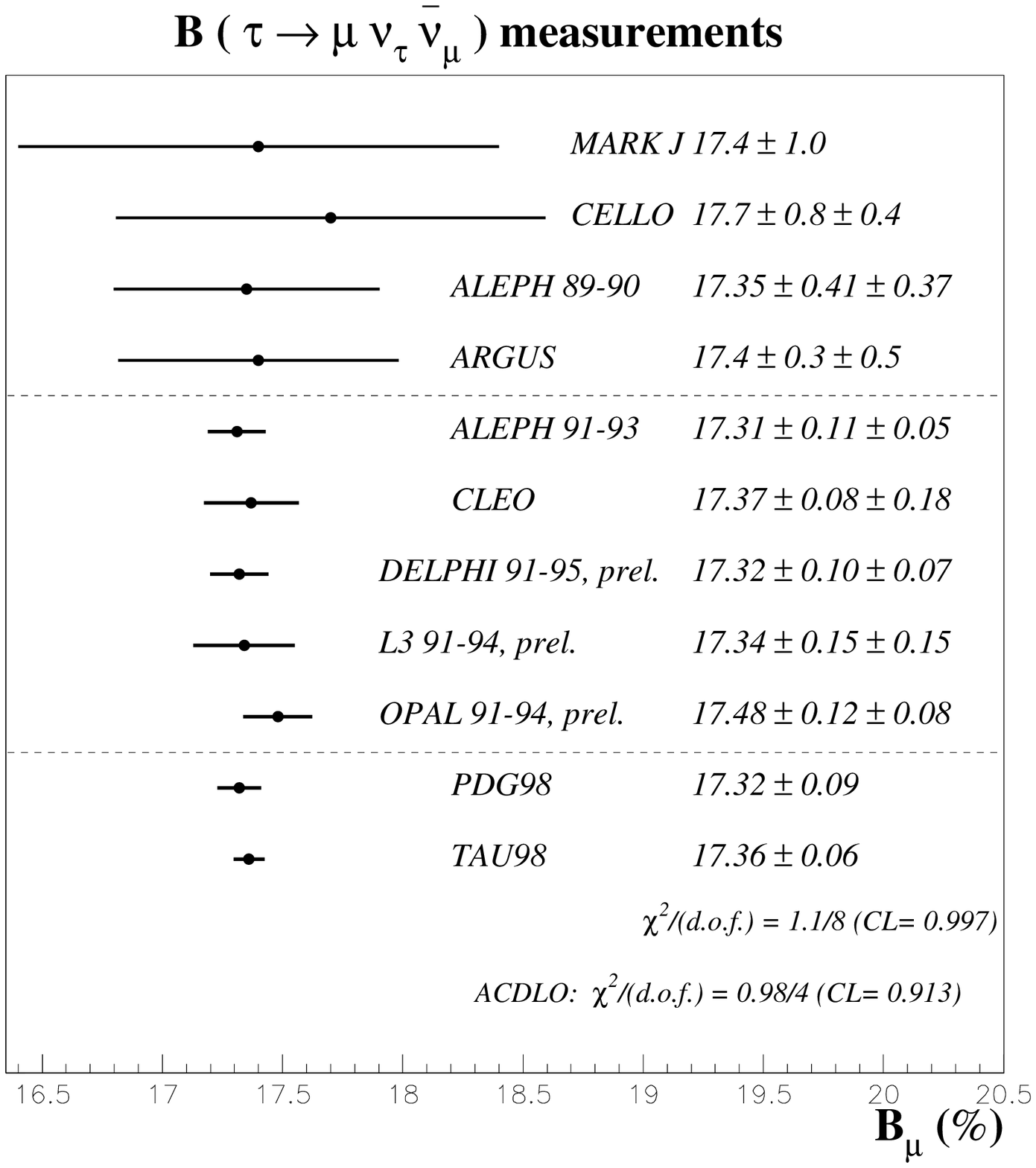}}
\end{center}
\vspace{-2.0cm}
\caption{ $B(\TMU)$ measurements for the world average.}
\end{figure}
The TAU98 world averages become:
$$
B(\TEL) = (17.81 \pm 0.06) \%
$$
and
$$
B(\TMU) = (17.36 \pm 0.06) \%.
$$
These branching ratios are thus now known
to a relative precision of 0.33\%, and are about 25\% more precise than the
PDG98 averages.
The $\chi^2$ confidence for the combination of the $B_{\mu}$, is
unnaturally high, about 99.7 \%. Disregarding psychological effects,
an explanation for
this can hardly be found, unless some measurements
have assigned too large values for the systematic uncertainty.
It may be noted that if one uses only the five most recent (and most precise)
measurements, the $\chi^2$ confidence is relatively normal,
at 91.3 \%. The level of agreement between the various
$B_e$ measurements is perfectly normal.
\par
To ensure the correct precision of the  $\e - \mu$ universality test
as given by eq. \ref{eq3}, 
it is necessary to account for  correlations in the measurements observed
by the different experiments, in particular in the CLEO measurements. 
The evaluation below therefore averages the ratios   $g_{\mu}/g_e$ when
they are given by experiments.  For the remaining measurements, 
averages of $B_e$ and $B_{\mu}$ are computed, and $g_{\mu}/g_e$ is
deduced from eq. \ref{eq3}. The following ratio is obtained:
$$
\frac{g_{\mu}}{g_{\e}} = 1.0014 \pm 0.0024
$$
This precision approaches the one obtained from pion decays, where
a comparison of the world average ratio:
$ R^{exp} = 
B(\pi \rightarrow \e \nu (\gamma))/B(\pi \rightarrow \mu \nu (\gamma)) =
(1.230 \pm 0.004) \times 10^{-4}$  \cite{PDG98} to
its theoretical predicition assuming universality , 
$R^{th} = (1.2352 \pm 0.0005) \times 10^{-4}$ \cite{sirlin2}, leads to
$$
 \left( \frac{g_{\mu}}{g_{\e}} \right)_{L} = 1.0021 \pm 0.0016
$$
Here, the subscript signals that this tests the coupling to a 
longitudinal $W$, and
is hence slightly different from the test from tau leptonic decays.
\par
To test $\tau$-e and $\tau$-$\mu$ universality, 
$\tau$ and muon masses and lifetimes are required. 
Using the TAU98 world average lifetime  $\tau_{\tau} = 290.5 \pm 1.0$ fs 
\cite{wasser}, and PDG98 numbers for the  other quantities, the
values
$$
\frac{g_{\tau}}{g_{\mu}} = 1.0002 \pm 0.0025
$$
and
$$
\frac{g_{\tau}}{g_{\rm{e}}} = 1.0013 \pm 0.0025
$$
are obtained. The uncertainties here 
are about equally shared between the uncertainty in the branching ratios
and the $\tau$ lifetime. 
\par
Finally, if e-$\mu$ universality is assumed, the
two branching ratios $B_e$ and $B_{\mu}$ can be combined to give:
$$
      B_l = (17.83 \pm 0.04) \%
$$
where the phase space suppression of $ B_{\mu} $ is corrected for.
This gives the ratio:
$$
\frac{g_{\tau}}{g_{\rm{e},\mu}} = 1.0007 \pm 0.0022
$$   
  Another test of $\tau-\mu$ universality can be derived 
by comparing $ B(\tau \rightarrow  h \nu)$ to  
$B( h \rightarrow \mu \nu)$.
Radiative corrections are now
known to a precision of a few permille in these ratios, both for $h=\pi$
and $h=K$ \cite{DECK1}, \cite{DECK2}. 
Uncertainties due to the hadron decay constant
and CKM element cancel when forming the ratio of these two numbers. 
As it is difficult to make the distinction between pions and 
kaons, CLEO \cite{CLEO} recasts the ratio by forming a ratio of   
$B_{h}=B(\tau \rightarrow  \pi \nu) + B(\tau \rightarrow  K \nu)$ and the sum
$H_{\pi} + H_{K}$. Here $H_{\pi}$ and $H_{K}$ are proportional
to $B(\pi \rightarrow \mu \nu)$ and $B ( K \rightarrow \mu \nu)$ 
respectively, given by
\begin{equation}
H_h = \frac{1+\delta_h}{\tau_h m_h} \left( 
      \frac{1-\frac{m_h^2}{m_{\tau}^2}}{1-\frac{m_{\mu}^2}{m_h^2}} \right)^2
B( h \rightarrow \mu \nu)
\end{equation}
with $h = \pi$ or $K$. The constants in front of the branching
ratios are choses such that the ratio
\begin{equation}
\frac{B_{\pi} + B_K}{H_{\pi} + H_{K}} = \frac{\tau_{\tau} m_{\tau}^3}
 {2 m_{\mu}^2} \left( \frac{g_{\tau}}{g_{\mu}} \right)^2 
\end{equation}
is independent of the pion and kaon decay constants ($f_{\pi (K)}$)
 and CKM elements.  With TAU98 average values for $B_h$ 
\cite{HELTSLEY} and the $\tau$  lifetime \cite{wasser} one obtains
$$
 \left( \frac{g_{\tau}}{g_{\mu}} \right)_L = 1.0037 \pm 0.0042
$$   
a precision which is  approaching the precision obtained
in leptonic $\tau$ decays. 
Again, the subscript signals that the spin structure of the coupling
is different here compared to the tests based on purely leptonic decays.
\par
Universality is also tested in a very direct way  through
the decay modes of real $W$ bosons \cite{WDEC1} \cite{WDEC2}. 
The sensitivity is presently about a factor of ten worse than in $\tau$ decays,
but will greatly improve as data come in from LEPII and future hadron
colliders.
\begin{figure}[t]
\begin{center}
\mbox{\epsfxsize 7.0cm \epsfbox{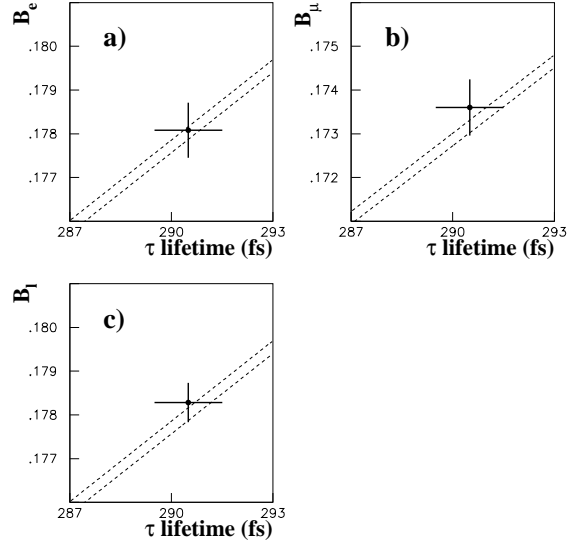}}
\end{center}
\vspace{-2cm}
\caption{ Branching ratios plotted versus the $\tau$ lifetime. a) 
$B_e$ b) $B_{\mu}$ , c) $B_{l}$, the combined leptonic branching ratio.
The region between the dashed lines  shows the expectation from 
equations 4  (fig. a) and c)) and 5 (fig. b), where the width of the
band is given by the uncertainty of the $\tau$ mass. }

\end{figure}
\section{ THE $\eta$ PARAMETER}
Departure from the standard model predictions for $B_e$ and $B_{\mu}$ does
not necessarily imply non-universal couplings to the
standard model $W^{\pm}$. The four fermion interaction
can be written in general form in terms of the Michel parameters,
and a non zero value of the $\eta$ parameter will affect the rates:
\begin{equation}
\label{eta}
 \Gamma_{l} =
 \frac{G_{l \tau}^{2}m_{\tau}^{5}}{192 \pi^3}\left[f(x_{l}^2)+
   4x_{l} g(x_{l}^2)  K \eta \right]r_{RC}.
\end{equation}
This expression is found in e.g. \cite{PICH}, but here a
factor $K$ is included to account for experimental
acceptance effects. $K=1$ when such effects are neglected.
Since the mass ratio $x_{\mu}$ is relatively large,
$ x_{\mu} = m_\mu / m_\tau \approx 1/17 $,  considerable sensitivity
is obtained by forming the ratio:
\begin{equation}
\label{eta2}
B_\mu/B_{\rm{e}} = 0.9726 +  0.217K \eta
\end{equation}
The parameter $\eta$ is the real part of the sum of
interference terms  between couplings of different Lorentz structure.
The sum includes the product between the standard model coupling and
the coupling of the $\tau$ to a charged scalar field. If an extra
Higgs doublet is present, there is a relation between $\eta$ and
the mass of the charged Higgs, given by \cite{hollik},\cite{stahl}: 
\begin{equation}
\eta = - \frac{m_{\tau} m_{\mu} \tan^2{\beta}}{2 m_H^2}
\end{equation}
where $\tan{\beta}$ is the ratio of the vacuum expectation values associated
to the two Higgs doublets. Other
contributions to $\eta$ have to be very small, as 
they would show up as the product of two non-standard couplings.
\par
$\eta$ is often called the low energy parameter, as
labaratory momentum spectra are mainly distorted at the low end for
non-zero $\eta$. This is just where experiments have
problems identifying muons, and the identification
efficiency drops to zero at some momentum cutoff, $p_c$.
Other experimental effects are probably not having a significant
impact on the sensitivity factor $K$, and a crude
estimate of the factor can be made based on the values of $p_c$
given in the different branching ratio analyses.
If the standard model shape is used to correct for
the number of events lost due to the cutoff, the 
sensitivity to $\eta$ is reduced by:
\begin{equation}
   K =           \frac{ \int_{p^c}^{p^{max}} h_{\eta}(p) dp} 
                { \int_{p^c}^{p^{max}} h_{s}(p) dp}
\end{equation}
where $h_s(p)$ is the standard model normalized momentum distribution and
$h_{\eta} (p)$ is the normalized distribution to multiply $\eta$.  
\par
For the branching ratio measurements, all four LEP experiments
make a momentum cut in the range 2 to 2.5 GeV/c, typically at
a momentum equal to 5 \% of the beam momentum.
In the CLEO analysis, all muon candidates are required
to have a momentum of at least $ 0.28 \times p_{beam} $  corresponding
to about 1.5 GeV/c. 
\par
\begin{figure}[t]
\label{ETA}
\begin{center}
\mbox{\epsfxsize 7.0cm \epsfbox{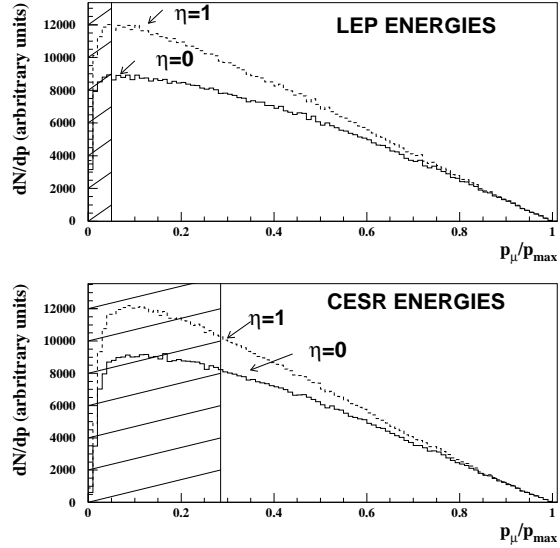}}
\end{center}
\vspace{-2cm}
\caption{  Laboratory momentum distribution of muons from $\TMU$
at LEP and CESR energies for the two hypotheses $\eta= 0$ and $\eta=1$. 
Muons with momenta in the hatched regions are
not identified by the experiments.}
\end{figure}

Generator level momentum distributions are shown in fig. \ref{ETA} and
based on these distributions the sensitivity factors are evaluated
at:
$K_{LEP} = 0.96$ and $K_{CLEO} = 0.72 $ 
Including this, eq. \ref{eta2} leads to the estimate:  
$$
 \eta = 0.012 \pm 0.024
$$
The assumption $K=1$ would have given a 10 \% smaller value
for the uncertainty.
\section{ CONCLUSIONS} 
Including the measurements presented at this conference, 
the averages for the $\tau$ leptonic branching ratios are:
$$
B(\TEL) = (17.81 \pm 0.06) \%
$$
$$
B(\TMU) = (17.36 \pm 0.06) \%
$$   
The present world averages tests e-$\mu$ universality to 
a precision of 0.24 \% at the level of the couplings. Furthermore, 
using lifetime and mass information, e-$\tau$ and $\mu - \tau$ universality
tests have a precision of 0.25 \%. The measurements also 
have implications for the Michel parameter $\eta$, and although some
care must be taken in evaluating the sensitivity, one can
conclude that $\eta$ is compatible with 
zero to a precision of 2.4 \%. No deviation from the standard model
predictions is found.
\par
The precision of these world averages  should improve by about
10 \% when all LEP data are fully analyzed. After this, little
improvement is expected in the foreseeable future, as CLEO and
the B factory measurements probably will be dominated by systematics
at the present level.   However, there should be
room for considerable improvements in the precision of ratios
between different branching fractions, for instance in $B_e/B_{\mu}$, with
corresponding improvement in $g_e/g_{\mu}$. 
\section*{ACKNOWLEDGEMENTS}
I would like to thank representatives for the different
experiments (B. Heltsley, W. Lohmann, R. Sobie and H. Videau) for
providing updated information on the current status of their 
$B_e$ and $B_{\mu}$ measurements,
and S. Wasserbaech for providing the updated $\tau$ lifetime estimate
before the start of the workshop. Finally I would like to thank
the organizers for a very instructive and enjoyable conference.

\end{document}